\documentstyle[12pt,epsfig]{article}
\newcommand{\scite}{\cite}
  \newcommand{\ccaption}[2]{
    \begin{center}
    \parbox{0.85\textwidth}{
      \caption[#1]{\small{{#2}}}
      }
    \end{center}
    }
\catcode`\@=11

\def\beq{\begin{equation}}
\def\eeq{\end{equation}}
\def\beqa{\begin{eqnarray}}
\def\eeqa{\end{eqnarray}}
\def\rd{{\mathrm d}}

\def\mev{\mbox{$\mathrm{MeV}$}}
\def\gev{\mbox{$\mathrm{GeV}$}}
\def\pt{\mbox{$p_T$}}
\def\jpsi{\mbox{$J\!/\!\psi$}}
\def\chic{\mbox{$\chi_c$}}
\def\psp {\mbox{$\psi'$}}
\def\ups {\mbox{$\Upsilon$}}

\def\height{\mbox{$H_8^{\prime}$}}
\def\vevpsi{\mbox{$\langle {\cal O}_8^{\psi}(^3S_1) \rangle$}}
\def\vevpsp{\mbox{$\langle {\cal O}_8^{\psi'}(^3S_1) \rangle$}}
\def\as{\mbox{$\alpha_s$}}

\newcommand{\bra}[1]{{\langle #1 |}}
\newcommand{\ket}[1]{{| #1 \rangle}}

\newcommand{\eqr}[1]{~(\ref{#1})}

\begin{document}

\begin{titlepage}

\begin{flushright}
\begin{tabular}{l}
            CERN-TH/95-129\\
            FNT/T-95/17 \\
	    IFUP-TH 25/95 \\
	    hep-ph/9505379 \\
            May   1995\\
\end{tabular}
\end{flushright}
\vfill

\begin{center}
{\huge Charmonium Production at the Tevatron}\\
\vfill
{\large Matteo Cacciari$^{a}$\footnote{Address after June 1 1995: DESY,
Hamburg, Germany}, Mario Greco$^{b}$,} \\
{\large Michelangelo L. Mangano$^{c}$ and Andrea Petrelli$^{d}$ } \\
\vspace{.5cm}
{\sl $^a$INFN and Dipartimento di Fisica Nucleare e Teorica, \\
Universit\`a di Pavia, Pavia, Italy \\
E-mail: cacciari@pv.infn.it }\\
\vspace{.2cm}
{\sl $^b$Dipartimento di Fisica,
Universit\`a di Roma III, Roma, Italy \\
INFN, Laboratori Nazionali di Frascati, Frascati, Italy \\
E-mail: greco@lnf.infn.it }\\
\vspace{.2cm}
{\sl $^c$CERN, Geneva, Switzerland \\
E-mail: mlm@vxcern.cern.ch }\\
\vspace{.2cm}
{\sl $^d$INFN and Dipartimento di Fisica,
Universit\`a di Pisa, Pisa, Italy }\\
\vfill
\begin{abstract}
We present in this work a study of large-\pt\ charmonium production in
hadronic collisions. We work in the framework of the factorization model of
Bodwin Braaten and Lepage, thereby including
the color octet production mechanism, and extract the values
of the necessary nonperturbative parameters from a comparison with the most
recent data from the Fermilab 1.8 TeV $p\bar p$ hadron collider. We extend the
calculation to 630 \gev, and compare the results with data published by the UA1
Collaboration. The global agreement is satisfactory, indicating that the
largest components of the production mechanisms for charmonium production at
high \pt\ have been isolated.
\end{abstract}
\end{center}
\vskip 1cm
CERN-TH/95-129\hfill \\
May 1995 \hfill
\vfill

\end{titlepage}

{\bf 1.}
The production of heavy quarkonium states in high energy processes has recently
attracted a lot of theoretical and experimental interest.
The successful operation of vertex  detectors in hadronic colliders has allowed
to disentangle the genuine charmonium yields from the large background due to
production and decay of $b$ quarks \cite{cdf94}.
The ability to detect the soft photons from \chic\ decays has
recently allowed the independent measurement of the \chic\ contribution to the
\jpsi\ rate \cite{cdfchi,d095,cdf95}.
On the theoretical side, charmonium production provides
quite stringent tests of our understanding of QCD on the very border between
perturbative and nonperturbative domains.
The detailed measurements of differential cross sections for
production of \jpsi, \psp\ and \chic\ states can be confronted with
theoretical calculations. Starting from the very intuitive Colour Singlet
Model \cite{csm}, these have recently been improved
with the inclusion of the mechanism of
production of a parton with large transverse momentum, followed  by the
fragmentation into charmonium states \cite{bygluon}.
The inclusion of the fragmentation mechanism has brought the theoretical
predictions closer to the observed prompt \footnote{Here and in the
following we use the term ``prompt'' to refer to all sources excluding
$b$-decay contributions.} \jpsi\ production rate \cite{frag}.
However the very large  discrepancy, by more than an order of magnitude,
between the theoretical  predictions and the data for the case of the \psp,
clearly demands for new mechanisms dominating the production process.

Several proposals have recently been put forward to solve this discrepancy.
Among these, the possible existence of    higher P-wave or D-wave states which
decay into the \psp,  or  of new metastable or hybrid charmonium states
\cite{close}.   In this paper we shall concentrate on a third proposal
\cite{bf}, namely the  contribution to the fragmentation function of colour
octet states, which subsequently evolve nonperturbatively into the \psp\  plus
soft light hadrons.

More recently, new data on the measurement of the \chic/$\psi$ fraction
\cite{d095,cdf95}\ confirm that a similar problem exists for \jpsi's not
coming form \chic\ decays.
The aim of the present paper is to make a thorough re-analysis of the full
matter, trying to find a coherent picture which possibly accounts for the
large \jpsi\ and \psp\ production cross sections, including the new information
available on the \chic\ production data.
The general framework is provided by the analysis of Bodwin, Braaten and Lepage
\cite{bbl}, which allows a consistent treatment of short and long distance
effects. We will first review the main ingredients of this formalism,
and then proceed to our phenomenological analysis.

{\bf 2.}
For the reader's convenience and to fix our notation, we briefly review
some models which have been suggested in connection with charmonium
production. We do not include a discussion of the color evaporation
model, for which phenomenological reviews have appeared recently
\cite{schuler}. We start instead presenting the Color Singlet Model (CSM)
\scite{csm} (for a recent review, see also \cite{schuler2}).
The dominant mechanism is assumed to be the short-distance
production of a color singlet $Q\bar Q$ pair
with the same spin and angular momentum quantum numbers of a given
quarkonium state $H$.
All the nonperturbative (long-distance) effects that lead to the formation
of the bound state
are factored into a single phenomenological parameter. Hence the cross section
for the production of a state $H=n\;^{2S+1}L_J$ takes the form
\beq
\sigma[n\;^{2S+1}L_J] = P_{nL}\,\sigma[Q\bar Q(n\;^{2S+1}L_J)],
\eeq
where the nonperturbative parameter $P_{nL}$ can then be expressed  in terms
of the radial wave function or its derivatives, and calculated either within
potential models (see for example ref.\cite{quigg}) or extracted from
experimental data.

This simple factorization fails in the calculation of  the production of $P$
states, for example via $q\bar q$ annihilation or in B mesons decays.  In fact
an infrared singularity appears, associated to a final state soft gluon, and at
least a second nonperturbative parameter has to be invoked to absorb it,
spoiling the simple minded picture of the CSM.

A rigorous framework  for treating quarkonium production and decays has been
recently developed by Bodwin, Braaten and Lepage \scite{bbl}.  Their so-called
``factorization model''
expresses the cross section for quarkonium production as a sum of terms each of
which contains a short-distance perturbative factor and a long-distance
nonperturbative matrix element, as
\beq
\sigma[H] = \sum_n {{F_n(\Lambda)}\over{m^{\delta_n-4}}} \langle 0|{\cal
O}^H_n(\Lambda)|0\rangle
\label{x-fact}
\eeq
$F_n$ are short-distance coefficients which can be calculated in  perturbative
QCD (pQCD) as power expansions in $\as(m)$ (the quark mass $m$ being large,
$\as(m)$ is expected to be small enough to allow the perturbative expansion).
$\Lambda$ is a scale which separates short and long distance
effects. The $\Lambda$ dependence of $F_n$ cancels
against that of the  matrix elements $\langle 0|{\cal
O}^H_n(\Lambda)|0\rangle$, leaving a cross section independent of $\Lambda$.
The above matrix elements can be rigorously  defined in Non Relativistic
QCD (NRQCD) \cite{bbl}.  They absorb the nonperturbative features of the
process and can either be extracted from data or calculated on the lattice.
Finally, the $\delta_n$ are related to the dimension of the operator ${\cal
O}^H_n$.

The main difference between the factorization model and the CSM is that not
all the operators
are now related to the production of color singlet $Q\bar Q$ pairs. The
factorization approach explicitly takes into account the complete structure of
the quarkonium Fock space. Therefore the quarkonium $H$ is no more
assumed to be simply a $Q\bar Q$ pair, but rather a superposition of states:
\beqa
|H=n\;^{2S+1}L_J\rangle&=&O(1)|Q\bar Q(n\;^{2S+1}L_J,\underline{1})
\rangle\nonumber\\
&+& O(v)|Q\bar Q(n\;^{2S+1}(L\pm 1)_{J'},\underline{8}) g\rangle\nonumber\\
&+& O(v^2)|Q\bar Q(n\;^{2S+1}L_J,\underline{8})gg\rangle + ... \; ,
\label{fock}
\eeqa
where the labels \underline{1} and \underline{8} refer to the colour state of
the   $Q\bar Q$ pair.
Higher order components are suppressed by powers of $v$, the average velocity
of
the heavy quark in the quarkonium rest frame. $v$ can be estimated through
the relation
\beq
v\simeq \as(mv)  \; ,
\eeq
which for the charmonium yields a value $v^2 \simeq 1/4$.
The CSM is recovered by taking only the lowest order term in
eq.\eqr{x-fact}.

The production of the state $H$ in the factorization approach can then proceed
via any of the Fock components in eq.\eqr{fock}. Higher order components
become important when their short distance coefficients $F_n$ are suppressed by
fewer powers of \as\
relative to lower order ones. Therefore the contribution of
various terms to the production of $H$ depends in general on both the
$\as(m)$ expansion of $F_n$ and the $v^2$ expansion of the matrix elements: it
is a two parameter problem.

Similar expressions hold within the factorization model
also for the fragmentation functions of a parton $k$ into the state $H$,
evaluated at a scale $\mu$
larger than the heavy quark mass:
\beq
D_k^H(z,\mu) = \sum_n {{d^k_n(z,\mu,\Lambda)}\over{m^{\delta_n-6}}}
\langle 0|{\cal O}^H_n(\Lambda)|0\rangle.
\eeq
The $d^k_n$ are, again, short distance coefficients. They can be
calculated in pQCD at a scale $\mu_0$ of the order of the quarkonium mass and
then evolved  to higher scales.
After evolution,  the cross section is
given by the usual convolution:
\beq
\sigma[H] = \int F^i F^j \sigma_{ij\to k} D_k^H
\eeq
the $F$'s being the parton distribution functions in the colliding hadrons
and $\sigma_{ij\to k}$ the kernel cross sections describing
the inclusive parton-parton scattering.

{\bf 3.}
Within
the factorization approach it is possible to relate the matrix elements of the
leading operator in the $v$ expansion to  those entering the factorization
formulae for quarkonium decays. They can therefore be extracted by comparing
the measured decay widths to those calculated. In the case of the operators
relative to higher components of the Fock space expansion, no simple relation
exists in general between decay and production matrix elements \cite{bbl}.   So
they should either be calculated (e.g. in lattice QCD), or
be measured directly in some production process.  We discuss here shortly
the cases of interest for our study. More detailed expressions and observations
can be found in \cite{bbl}.

In the case of \chic\ production, we have the following
expressions for cross sections and
fragmentation functions to leading order in $v^2$:
\beqa
&&\sigma[\chi_J]
= {{F_1(^3P_J)}\over{m^4}} \;
	\bra{0} {\cal O}_1^{\chi_J}(^3P_J) \ket{0}
\;+\; {F_8(^3S_1) \over m^2} \;
	\bra{0} {\cal O}_8^{\chi_J}(^3S_1) \ket{0}
\quad\quad J = 0,1,2\\
&&D_g^{\chi_J}
= {d^g_1(^3P_J) \over m^2} \;
	\bra{0} {\cal O}_1^{\chi_J}(^3P_J) \ket{0}
\;+\; d^g_8(^3S_1)\;
	\bra{0} {\cal O}_8^{\chi_J}(^3S_1) \ket{0}
\quad\quad J = 0,1,2
\eeqa
The presence of the color octet matrix elements represents the  natural
extension of the CSM results, and allows the absorption of the infrared
divergences which appear in the short distance coefficients of the color
singlet part. Both terms are needed to give a consistent description of \chic\
production at this order in $v^2$.

The matrix elements of color singlet operators can be
related to those entering the decay processes
$\chi_{J=0,2}\to \gamma\gamma$ and $\chi_{J=0,2}\to$ light hadrons.
The color octet production matrix element, however,
cannot be related to the corresponding decay one \cite{bbl92}.
We have to resort to a production process to
measure it. In \cite{bbl92}\ it was suggested to
use the  \chic\ production in $B$ decays.
The results have been reported in the
literature in terms of the nonperturbative parameters $H_1$ and $H_8'$ for
the color singlet and color octet parts respectively. They are related
to the NRQCD matrix element as follows:
\beqa
&& H_1 = {1\over{m^4}}{{\bra{0} {\cal O}_1^{\chi_J}(^3P_J)
\ket{0}}\over{2J+1}}\\
&& H_8' = {1\over{m^2}} {{\bra{0} {\cal O}_8^{\chi_J}(^3S_1)
\ket{0}}\over{2J+1}}
\eeqa
In ref.\cite{bbl92} $H_1$ was
obtained fitting the $\Gamma(\chi \to \rm{light\ hadrons})$, whereas in
ref.\cite{bf} $H_8'$ was extracted from the CLEO measurement of
$BR(B\to\chi_{J=1,2} + X)$ \scite{cleo}. The values found are:
\beqa
&& H_1 \approx 15~\mev \\
&& H_8' = 1.4 \pm 0.6~\mev \label{cleo}
\eeqa

Let us now consider $^3S_1$ states, i.e. $\jpsi$ and $\psi'$. We
will collectively indicate these as $\psi$.
The cross section and fragmentation function for producing a $\psi$ to leading
order in $v^2$ are simply given by the CSM results:
\beqa
\sigma[\psi] &=&
      {{F_1(^3S_1)}\over{m^2}}\langle 0|{\cal O}_1^\psi(^3S_1)|0\rangle \\
D_k^\psi &=&
      d^k_1(^3S_1)\langle 0|{\cal O}_1^\psi(^3S_1)|0\rangle
\eeqa
The matrix element appearing in the above equations can be shown to be related
to the standard nonrelativistic wave function $\overline{R_\psi}$
as follows \cite{bbl}:
\beq
\langle 0|{\cal O}_1^\psi(^3S_1)|0\rangle
\simeq
{9\over{2\pi}} \Big|\overline{R_\psi}\Big|^2
\eeq
They can therefore be extracted from the measurement of the
leptonic decay width of the $^3S_1$ states, or can be calculated within
potential models \cite{quigg}.

While no color octet contribution appears at order $v^2$, it
has recently been argued by Braaten and Fleming \scite{bf}\ that yet higher
order terms can however be significantly enhanced since their short distance
coefficient appears at lower
orders in the $\as$ expansion.
An example of this is gluon
fragmentation to $\psi$. To leading order in $v^2$ the  fragmentation
proceeds through a color singlet $^3S_1$ state, and starts to order $\as^3$:
\beq
D_g^\psi = \as^3 \hat D_1 \bra{0} {\cal O}_1^{\psi}(^3S_1) \ket{0}
\label{g-singlet}
\eeq
This is because production of the $c\bar c$ pair in a color singlet state
requires emission of two perturbative gluons.
On the other hand, the fragmentation process where the gluon goes into a
color octet state, although suppressed by $v^4$, starts at order $\as$:
\beq
D_g^\psi = \as \hat D_8 \bra{0} {\cal O}_8^{\psi}(^3S_1) \ket{0}
\eeq
It can therefore be numerically relevant when compared to \eqr{g-singlet}.
No decay process is known which is dominated by the color octet component, and
therefore it is not possible to extract the relative matrix elements from
decay widths. One could get a crude estimate of their values by rescaling the
color singlet matrix elements by the appropriate powers of $v$.
We prefer here to take their value as a free parameter, to be fitted to the
Tevatron production data. We will verify at the end that the results are
consistent with the expected $v^4$ suppression.

{\bf 4.}
We now present a comparison between experimental results
and the calculations illustrated above. We first concentrate on results from
the Tevatron collider, relative to high-\pt\ production of \jpsi, \psp\
and \chic\ states. Since we will present results for prompt production,
we will only make use of the CDF data, for which the $b$-decay background has
been removed \cite{cdf95}. It is important
to stress, nevertheless, that there is perfect agreement between the CDF and
the D0 data
when all sources of \jpsi\ are included \cite{d095}.
We will extract the values of the nonperturbative parameters \height,
\vevpsi\ and \vevpsp\
from fits to the experimental data. We will
then use these values to ``predict'' the inclusive \jpsi\ \pt\ distributions at
the energy of 630 \gev, where data are available from the
measurements performed by the UA1 experiment \cite{UA1psi}.

For the calculation of large $\pt$ charmonium production in $p\bar p$
collisions at the Tevatron energy ($\sqrt{s} = 1800$~GeV) we include the
following contributions:
\begin{enumerate}
\item Direct production of charmonium states. The matrix elements were
calculated in
Ref.~\cite{csm}. As previously noted in literature \cite{frag}, this
contribution is very small compared to the fragmentation one.
\item Production via fragmentation of gluons and charm quarks. These
contributions were considered in ref.\cite{frag}, where it was shown that they
greatly enhance the cross sections with respect to the direct terms.
We use the fragmentation functions of gluon
to $\psi$ \scite{bygluon}, charm to $\psi$ \scite{bcycharm} and gluon to
$\chi$ \scite{bychi}.
\item Production via fragmentation into color octet states.
\end{enumerate}
\begin{figure}[t]
\begin{center}
\psfig{figure=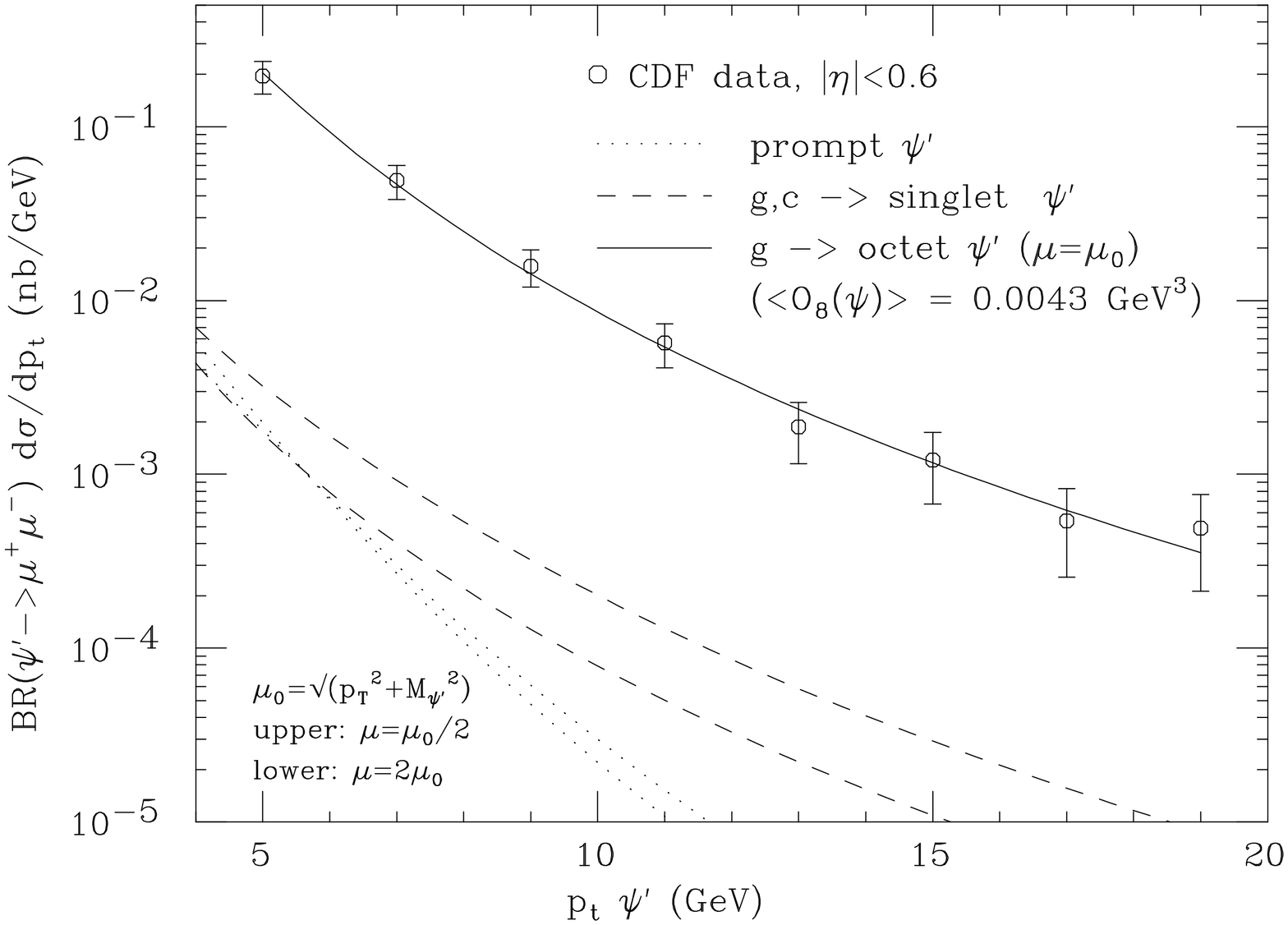,width=10cm,angle=90,clip=}
\ccaption{}{\label{fig-psp}\small Inclusive prompt $\psi'$ $p_T$ distribution.
CDF data versus theory. The contribution from different sources is shown.}
\end{center}
\end{figure}
\begin{figure}[t]
\begin{center}
\psfig{figure=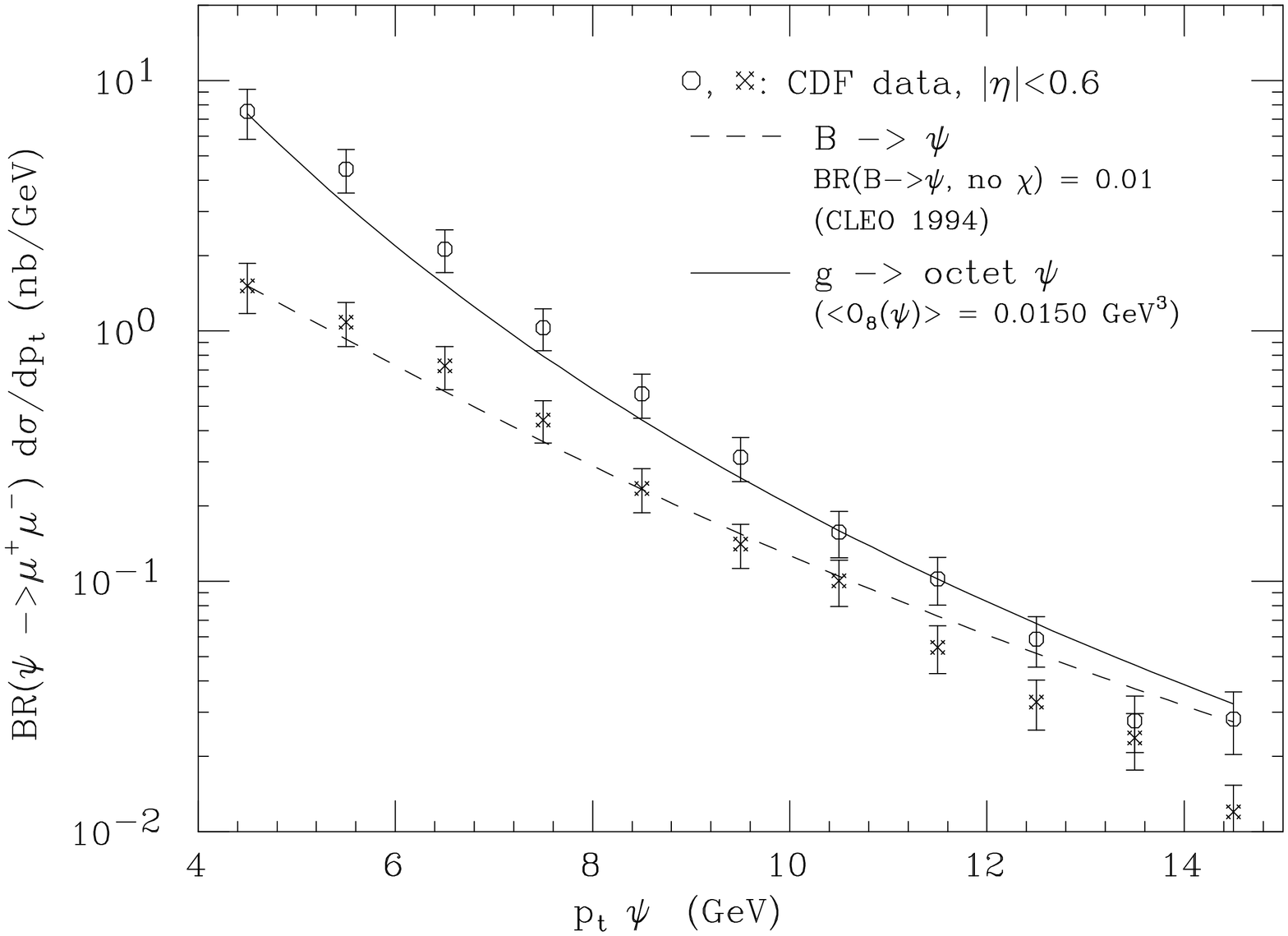,width=10cm,angle=90,clip=}
\ccaption{}{
\label{fig-psi}\small Inclusive $\psi$ $p_T$ distribution.
Upper curves and data points correspond to prompt $\psi$'s, after subtraction
of the $\chi_c$ contribution. Lower ones correspond to the $b$ decay
contribution.
CDF data versus theory.}
\end{center}
\end{figure}
We will also show separately the contributions to \jpsi\ and \chic\
production coming from the decay of $b$ quarks,
evaluated at the next-to-leading order \cite{nde}\ using a choice of
renormalization and factorization scales which provides the best fit to the
Tevatron data \cite{fmnr}.

All the charmonium cross sections are evaluated at leading order with the MRSA
\scite{mrsa} parton distribution set. The renormalization/factorization scale
is set at $\mu = \sqrt{\pt^2 + M_\psi^2}$. We have checked that using for $\mu$
the \pt\ of the fragmenting parton produces differences which are typically of
order 10-20\%, therefore definitely less than the other uncertainties involved.

Fig.~\ref{fig-psp}\ shows the comparison between theory and CDF
data \scite{cdf95} for prompt \psp\ production.
In this and in the following figures we have not shown the
band due to theoretical uncertainties due to, for example, the choice of the
renormalization, factorization and fragmentation scales \cite{frag}.
This is because these uncertainties mostly affect the overall
normalization of the curves, and not their shape. As a consequence, their
effect would be hidden by a rescaling of the fitted value of the
nonperturbative parameters.

As was already shown in the work of Braaten and Fleming \cite{bf},
the theoretical curve agrees well with the shape of the data.
The old theoretical prediction from pure color singlet fragmentation
was known to fall a factor of 30 below the CDF data, as shown in the figure.
The addition of the color octet mechanism  reconciles theory and data. The
value we extract for \vevpsp\ from a best $\chi^2$ fit is $4.3 \times 10^{-3}$
$\gev^3$, close to what derived in \cite{bf}.

\begin{figure}[t]
\begin{center}
\psfig{figure=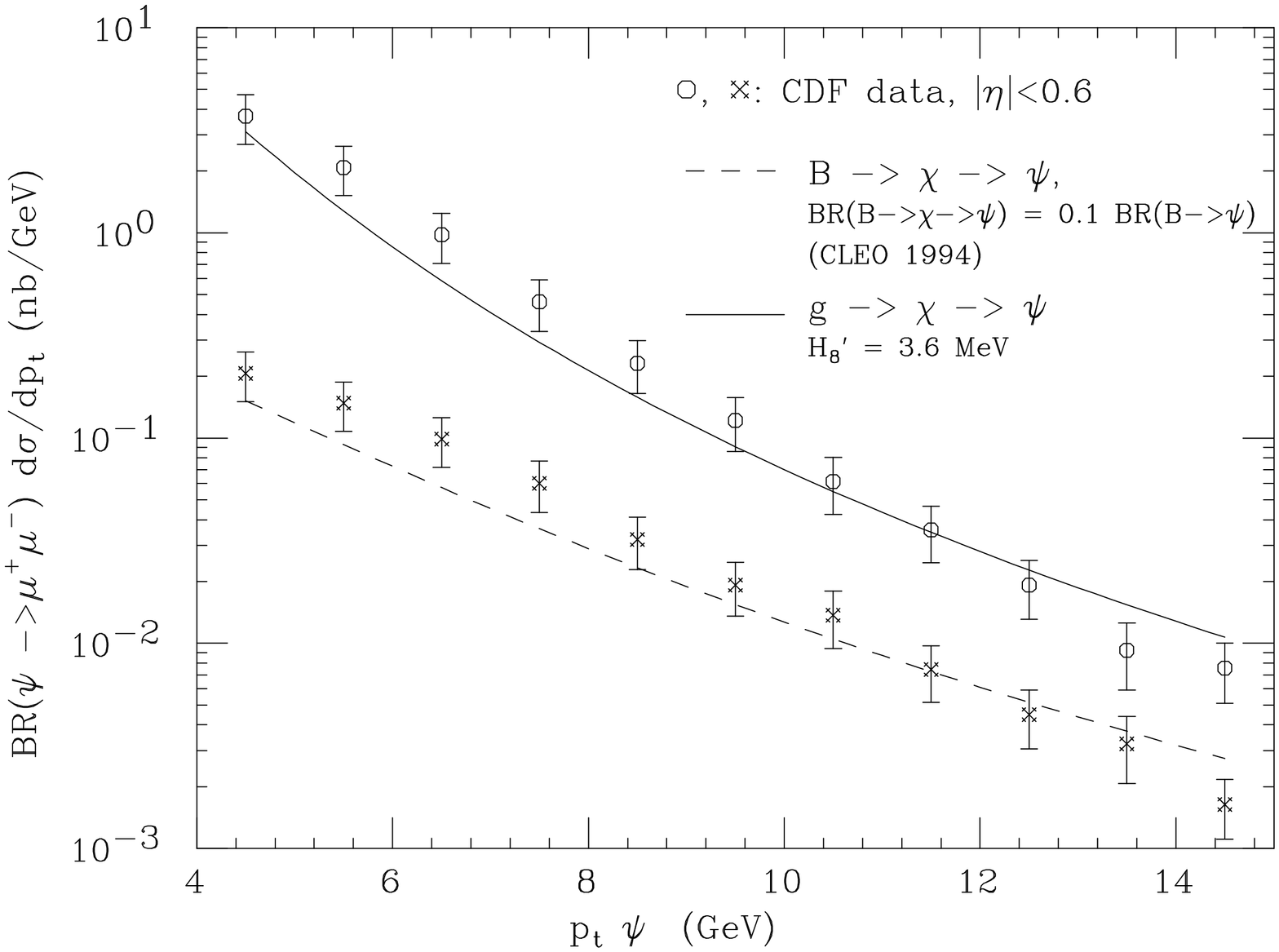,width=10cm,angle=90,clip=}
\ccaption{}{
\label{fig-chi}\small Inclusive $p_T$ distribution of $\psi$'s from
$\chi_c$ production and decay.
Upper curves and data points correspond to the prompt component.
Lower ones correspond to the $b$ decay contribution.
CDF data versus theory.}
\end{center}
\end{figure}

Fig.~\ref{fig-psi}\ shows the inclusive \pt\ distributions of \jpsi\
{\em not} coming from \chic\ decays. Both prompt and $b$-decay contributions
are shown separately.
The value we extract for \vevpsi\ from
a best $\chi^2$ fit to the prompt data is $15 \times 10^{-3}$ $\gev^3$.
Without inclusion of the color octet components, the disagreement between
theory and data would be similar that that previously noted for \psp's, namely
a factor of the order of 30.

The
ratio \vevpsi/\vevpsp\ is about 3, which is consistent with
the ratio of the values of the color singlet wave functions at the
origin. In other words, the values one extracts from the two independent sets
of data are not unnatural within the color octet scheme. An intrinsic
uncertainty in the comparison of these numbers comes from the ambiguity present
in the choice of the mass scales. For example, using $M_\psi$ or $M_{\psi'}$
as opposed to $2m_c$ in the coefficient function, can lead to variations up
to a factor of 2 in the fit results, and in their ratios.
As for the absolute value of the matrix elements, these are consistent with a
suppression of the order of $v^4 \simeq 0.06$ relative to the color singlet
ones.

\begin{figure}[t]
\begin{center}
\psfig{figure=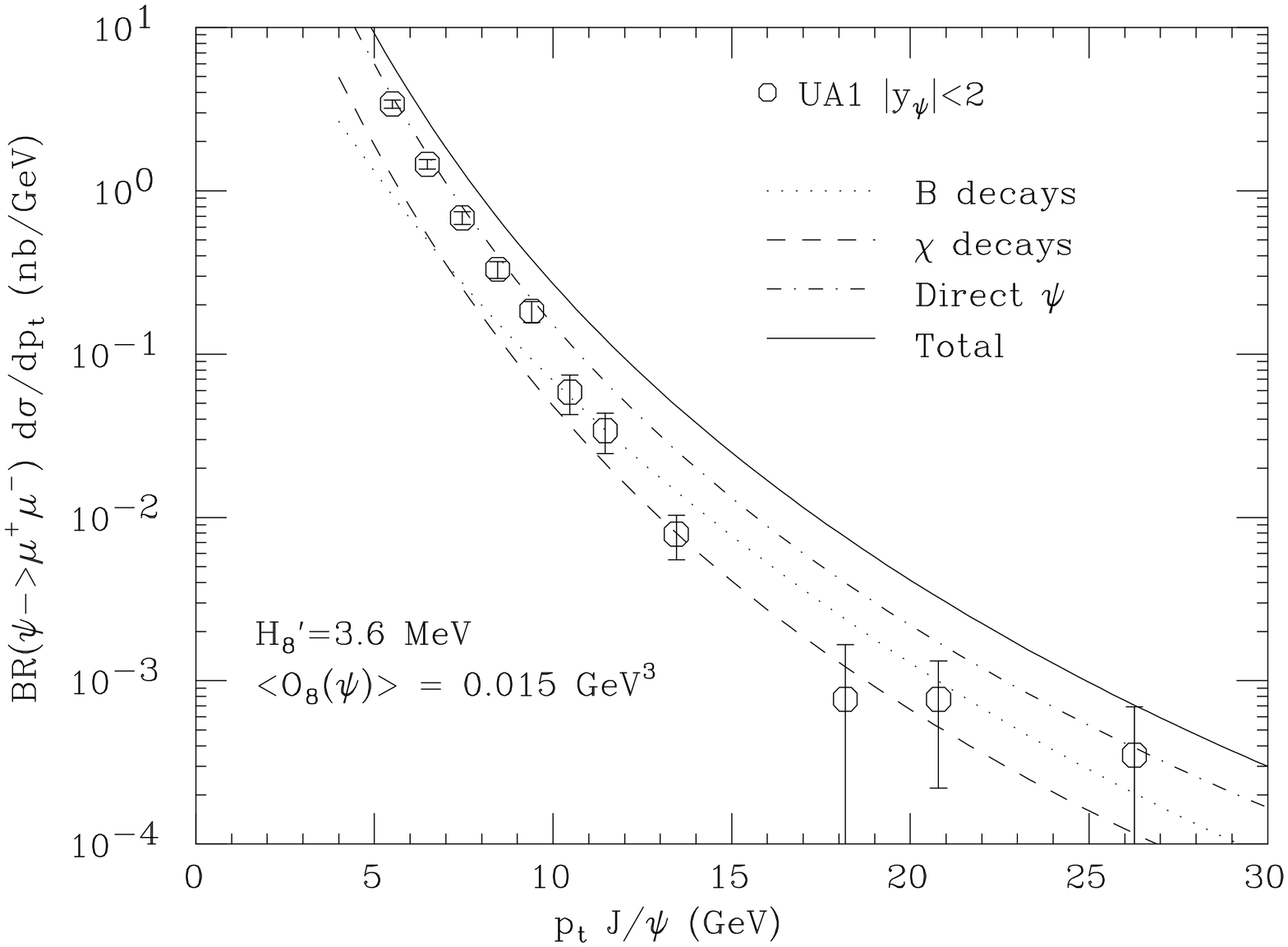,width=10cm,angle=90,clip=}
\ccaption{}{
\label{fig-UA1}\small Inclusive $p_T$ distribution of $\psi$'s at 630
GeV. All sources of $\psi$ production are here included.
UA1 data versus theory. The parameters of the
theoretical calculation take the values fitted on the Tevatron data. }
\end{center}
\end{figure}

A precise prediction of the color octet mechanism, however, is that for
sufficiently large \pt\ the ratio of the \jpsi\ and \psp\ cross sections should
be a constant. Current data do not fully support this expectation.
It is in fact possible to obtain a good fit to the
data shown in Figs.~\ref{fig-psp}\ and \ref{fig-psi}, in the common range
$4<\pt<15$ \gev, using the following parametrizations \cite{mlmfit}:
\beqa
{{\rd\sigma(\jpsi)}\over{\rd\pt}} = 1773 \exp(-\pt/1.65) \\
{{\rd\sigma(\psi')}\over{\rd\pt}} = 384 \exp(-\pt/1.79)
\eeqa
These fits predict a ratio which is rising with $\pt$,
varying from 0.2 to 0.4 (after removing
the BR's) in the \pt\ range currently accessible.
Taking into account the experimental uncertainties, this is
not inconsistent with a constant ratio. If however future improved statistics
should confirm this trend, this would be a clear indication that yet more
mechanisms, such as multiple decays of higher charmonium resonances, are at
work.

To conclude the survey of charmonium production at the Tevatron,
we present in Fig.~\ref{fig-chi}\ the \pt\ distribution of \jpsi's coming
from \chic\ decays. The theoretical curves include the effect of \pt\ smearing
due to the $\chic\to\jpsi$ decays.
Both theory and data use the recent determination of $BR(b\to\chic_J+X)$
from CLEO \cite{cleo}\ to extract
the $b\to\chic\to\psi \, \gamma$ contribution.

The best $\chi^2$ fit to the prompt data gives a value of \height=3.6 \mev.
The shapes of theory and data are consistent with each other, although the
agreement is not as good as in the case of \psp.
The value of \height\ is larger by a factor of 2 relative
to that measured by CLEO using
$b\to\chic$ decays, Eq.~\ref{cleo}.
It should be kept in mind that the values extracted from
the fits to the hadron collider data are directly sensitive to the perturbative
$K$ factors due to higher order corrections to the hard process matrix element,
and to the fragmentation functions.
As a reference, the NLO $K$ factor for the production of a large \pt\ gluon
was evaluated to be approximately 1.5 in the work of Cacciari and
Greco, Ref.~\cite{frag}.

Having fixed the values of the nonperturbative parameters using the
Tevatron data, it is possible to use them to make predictions for different
energies and different beam types.
We consider here data published by the UA1 Collaboration
\cite{UA1psi}, relative to interactions at the 630 \gev\ CERN $S\bar p pS$
Collider. We expect that at this energy and at the \pt\ values measured by UA1
the production mechanisms should be exactly the same as those active at the
Tevatron. This would be true even in presence of additional processes, such as
for example production and decay of higher charmonium resonances.

We present these data in Fig.~\ref{fig-UA1}, together with the theoretical
predictions. UA1 measured the \jpsi\ \pt\ spectrum inclusive of all
contributions from $b$- and \chic-decays. The contribution of \psp\ decays,
once convoluted with $BR(\psp\to\jpsi)$ and with the decay spectrum, amounts
to much less than 10\% of the total, and was neglected here. The comparison
shows that theory predicts now approximately a factor of two more
\jpsi's than are observed. In view of what was said few lines above, this is
contrary to our expectations. We only see one possible explanation for this
discrepancy, leaving out experimental systematics. Namely the significant
difference in perturbative $K$ factors at the two energies. It has been
observed
since a long time that $b$ production cross sections at 1.8 TeV are
systematically higher than theory. Agreement with NLO QCD can be found only by
choosing extreme values of the renormalization scale, or choosing values of
$\alpha_s$ larger than the input parton distribution sets prescribe, in
addition to fixing the $b$ mass to the relatively low value of
4.5 \gev \cite{fmnr}.
By choosing as input parameters for the theoretical evaluation of the $b$ cross
section at 630 \gev\ the same values that fit the normalization of the Tevatron
data, one finds a result which is approximately 30-40\% higher than the UA1
data \cite{mlm-glasgow}. A justification for such a discrepancy can be found in
the study of small-$x$ effects in heavy quark production at high energy
\cite{smallx}. It is expected that such effects should be larger for production
of $c$ quarks, although no detailed estimate exists. If this were indeed the
case, however, the relative discrepancy of a factor of 2 between charmonium
production at the 1800 and at 630 \gev, as found without inclusion of small-$x$
effects, would be consistent with the similar discrepancy by a factor of
30-40\% found in the case of $b$ production.

{\bf 5.}
We considered in this paper the large \pt\ production of \jpsi, \psp\ and
\chic\ states via gluon fragmentation into the leading color singlet and
color octet components of their wave function.  The result
of a previous study by Braaten and Fleming of \psp\ production extends to the
case of the \jpsi, showing that these effects can explain the unexpectedly
large rate of prompt \jpsi\  and the small \chic/$\psi$ production ratio
observed at the Tevatron. The values of the nonperturbative parameters needed
to parametrize these production processes turn out to be consistent with what
naively expected.

The extension of these calculations to the case of inclusive \jpsi\ production
at 630 \gev\ results in rates which are approximately a factor of 2 larger than
measured by UA1. We attribute this discrepancy to a larger $K$ factor at the
higher energy, due to more important small-$x$ effects.

Current data from the Tevatron and the residual theoretical uncertainties
cannot exclude the presence of yet additional production mechanisms, such as
the production and decay of higher resonances. Dominance of the production via
fragmentation into the color octet component of the $^3S_1$ states strictly
predicts \psp/\jpsi\ to be a constant, at least for $\pt\gg M_\psi$.
Current data do not support this conclusions, although the statistical
uncertainty is still large.

This formalism cannot be directly applied to the calculation of total cross
sections, or to the region $\pt<M_\psi$. This is because in this region the
fragmentation approximation is not justified. The effect of color octet
production, however, can be calculated including the full set of relevant
Feynman diagrams. After this work was completed, we received a paper by Cho and
Leibovich \cite{cho}\ in which this calculation has been performed, and applied
to charmonium and small-\pt\ \ups\ production at the Tevatron. The results of
their work are consistent with ours over most of the \pt\ range covered by the
charmonium data. Their value of \height\ is smaller than what we find,
presumably because of the  the absence in their calculation of the {\em
negative} color singlet contribution to \chic\ production \cite{bychi}. Their
value of \vevpsp\ is larger than ours, presumably because their calculation
correctly incorporates the small \pt\ decrease in rate relative to the
fragmentation approximation.
The values of the nonperturbative parameters extracted from the fits to the
Tevatron data can then be used, in association to the matrix elements evaluated
in \cite{cho}, to perform more precise predictions of total cross sections at
fixed target energies, where a large amount of data is available.

\end{document}